\documentclass{aa}

\newcommand{\fwhm}{{\sc fwhm}}
\newcommand{\code}{\texttt} 
\newcommand{\rii}{{\em right}} 
\newcommand{\lee}{{\em left}} 
\newcommand{\isis}{{\sc isis}} 
 
\newcommand{\daophot}{{\sc daophot}} 
\newcommand{\allstar}{{\sc allstar}} 
\newcommand{\allframe}{{\sc allframe}} 
 
\newcommand{\idl}{{\sc idl}}

\newcommand{\mmag}{\,mmag}

\newcommand{\vs}{{\em vs.}}

\newcommand{\sperl}{{\sc period98}\index{period~ninety~eight@{\sc period98}}}

\newcommand{\ea}{et~al.}
\newcommand{\ie}{i.e.} 
\newcommand{\eg}{e.g.}

\newcommand{\cf}{{cf.}}

\newcommand{\Sec}{${}^{\prime\prime}$\llap{.}}
\newcommand{\secx}{${}^{s}$\llap{.}}
\newcommand{\Secc}{${}^{\prime}$}
\newcommand{\teff}{$T_{\rm eff}$}

\newcommand{\ocen}{$\omega$~Cen\index{Omega Cen@{$\Omega$ Cen}}}
\newcommand{\ngcsyv}{NGC~6791} 
\newcommand{\cfht}{{\code{CFHT}}\index{cfht@{\code{CFHT}}}}   
\newcommand{\alfosc}{{\code{ALFOSC}}\index{alfosc@{\code{ALFOSC}}}}   

\usepackage{graphicx}

\begin{document}

   \title{A Search for Planets in the Old Open Cluster NGC~6791
          \thanks{Based on observations made with the 
Nordic Optical Telescope,
operated on the island of La Palma jointly by Denmark, Finland,
Iceland, Norway, and Sweden, in the Spanish Observatorio del 
Roque de los Muchachos of the Instituto de Astrof\'isica de 
Canarias.}}

    



   \author{H. Bruntt\inst{1}
          \and F.\ Grundahl\inst{1}
          \and B.\ Tingley\inst{1} 
          \and S.\ Frandsen\inst{1}
          \and P.\ B.\ Stetson\inst{2}
          \and B.\ Thomsen\inst{1}
          }
   \offprints{H.\ Bruntt, bruntt@phys.au.dk}

   \institute{Department of Physics and Astronomy, University of Aarhus,
              Ny Munkegade, 8000 Aarhus C, Denmark
         \and 
              Dominion Astrophysical Observatory, Herzberg Institute of
              Astrophysics, National Research Council, 5071 West Saanich Road, Victoria,
              British Columbia V9E 2E7, Canada
             }

   \date{Received ; accepted}

   \abstract{ We describe the results of a search for 
              transit-like events caused by giant planets occulting
              stars in the old, metal-rich open cluster
              NGC~6791 based on $BVI$-colour photometry from eight nights of 
              observations with the
              2.54\,m Nordic Optical Telescope. To extract the light curves 
              we have used both PSF photometry (\daophot) and the difference 
              imaging technique (\isis). We have re-analyzed
              observations from earlier campaigns to search for multiple
              transits, determine periods of long-period 
              variables, and detect eclipsing binaries.  
              We confirm 20 known variables and have discovered 22 new low-amplitude 
              variables with amplitudes in the range 7--40\mmag\ and periods
              1--16 days. We have found the primary and secondary 
              eclipses of two eclipsing binaries based on the new and
              older photometric campaigns. The search for transits-like events has turned up
              a few single-transit candidates. The transit depths are
              10\mmag\ in both the $V$ and $I$ filters over periods of 1.0--2.5 hours, 
              but future observations are required to see whether identical
              transit-like events recur in these same stars. 
   } 

   \maketitle

\section{Introduction}

  Since the discovery of the first extra-Solar giant planet 
  around 51 Peg (Mayor \& Queloz \cite{mayor95}), around
  100 extra-Solar planets have been discovered. 
  Nearly all of these have been found by measuring the variation in 
  radial velocity caused by the orbit of the planet
  around the parent star. 
  In the case of HD~209458, the planet is known to 
  transit the star causing a reduction of 1.7\% in the light from the star.
  From the photometric and radial velocity curves 
  Charbonneau \ea\ (\cite{charbonneau00}) could determine the mass 
  and radius of the planet to within a few percent.
  Interestingly, the resulting mean density of the planet is quite low due the
  close vicinity of the parent star (Burrows \ea\ \cite{burrows2000}).

Several attempts have been made to
detect extra-Solar planets by monitoring many thousand stars 
simultaneously with an imaging CCD camera in order to detect the reduction
in light due to the transit of a planet\-ary companion. Most notable is the
result for 47~Tuc (Gilliland \ea\ 2000), which failed to detect planets
during an 8.3~day search with the Hubble Space Telescope. 


In addition, Udalski \ea\ (2002a) selected a sample of 52\,000 stars 
from the OGLE-III (Wozniak \ea\ 2002) campaign. 
These stars have photometry better than 
1.5\% and were observed over 32 nights during a period of 45 days. 
Udalski \ea\ reported 47 transiting systems of 
which 42 exhibited multiple transits.
An additional 13 objects were later reported by Udalski \ea\ (2002b), who used
a refined search method.  
In a subsequent paper Dreizler \ea\  (2002) presented spectra of 16 of 
the most promising planet transit candidates found by Udalski \ea\ (2002a).
From these spectra Dreizler \ea\  (2002) could determine 
the spectral classes of the stars and 
in turn constrain the mass and radius ratios of the 
orbiting object and the parent star. Dreizler \ea\  found that
only two of the 16 transiting objects have radii and masses consistent 
with being giant planets.
The other transiting objects are probably brown dwarfs or M-type stars. 

Gonzalez \ea\ (2000) have made spectroscopic studies to determine the
metallicities of stars with planets and from a sample of 39 stars they find
$<$[Fe/H]$>=0.18\pm0.19$. In comparison, the local average metallicity of 
G-type dwarfs is [Fe/H]~= $-0.2$~dex (Rocha-Pinto \& Maciel 1996).  
Laughlin (2002) states that the probability of a star hosting a
planet is 10 times higher for stars with high metallicity ([Fe/H] $>0.2$)
than for less metal-rich objects.  One possible reason for
the higher metallicity of stars with planets could be that planets are 
more likely to form in a metal-rich environment (see Laughlin 2002 for 
a discussion).  Another possibility (Laughlin 2002) could be the addition 
of metal-rich material onto the star, which is mixed up in the convective 
outer layers.  
The source could be migrating proto-planets and/or 
planetesimals from the proto-planetary disc.

Laughlin (2000) also examined the correlation between 
metallicity and mass of stars with giant planets. Among stars with
planets there is a tendency for the stars with higher metallicity to also
have larger mass.  The proposed explanation is the fact that the higher the 
mass the shallower the outer convection zone will be, hence for the more 
massive stars the metal-rich material which is added is mixed through a smaller 
fraction of the star. 

Inspired by these interesting results we selected the old {\em metal-rich}
open cluster NGC~6791 as 
our target for a search for planet transits. The metallicity of the cluster
is a subject of debate. Taylor (2001) states that ``the best that can be said 
at present is that the cluster metallicity lies in the range from +0.16 to 
+0.44 dex''.  Two studies published since the Taylor article come to 
no better agreement:  Friel \ea\ (2002) have derived [Fe/H] = $+0.11\pm0.010$
from spectroscopy of cluster giants, while Stetson \ea\ (\cite{stetson03}) infer
an abundance $\sim\,$+0.3 by matching the morphology of the
cluster color-magnitude diagram to that of recent theoretical isochrones.

In comparison the globular cluster 47~Tuc studied by Gilliland \ea\  (1999) has
[Fe/H] = $-0.7$ (Salaris \& Weiss 1998).
In this study around 30 transits of short-period
giants were expected from the 34\,000 light curves, i.e.\ one out of every 1\,000 stars.
The fact that no transits were found is perhaps due to the 
lower probability of finding planets in a metal-poor system. Another
possible explanation is that the relatively high stellar density in a globular cluster
may have influenced the evolution of proto-planetary discs or the
long-term stability of any planetary systems that may have formed.
The dynamical properties of open clusters are not as extreme as 
for globular clusters and the formation of planets may take place here.

We observed \ngcsyv\ for $\simeq8$~hours on each of 
seven nights and we estimate the probability of detecting a single 
transit of a close-in planet to be $P_{\rm single} \simeq\ 50\%$ 
and $P_{\rm twice} \simeq 7\%$ for detecting a transit twice for the same star.
The probability that a local G or F-type field star harbours a close-in 
giant planet is around 0.7\,\% (Butler \ea\ \cite{butler02}) while the 
geometric probability that 
the planet will cause a transit is 10\,\%.
We have $N_{\sigma < 2\%} = 2\,500$ stars in our sample 
with light curves with a precision better than 2\% per data point. 
Assuming that the binary fraction $P_{\rm bin}$ of the 
cluster is 15\% the expected number of planets that will be detected is
\begin{equation}
N = 
N_{\sigma < 1\%} * P_{\rm GF} * P_{\rm geom} * (1-P_{\rm bin}) * P_{\rm single} = 0.8
\end{equation}


The metallicity of
\ngcsyv\ is significantly higher than the field stars and we may expect 
to find at least a few candidates with single transits.  
According to Laughlin (\cite{laughlin00}) the fraction of stars harbouring planets 
is $\simeq$\,10 times higher for metal rich stars.

If we detect transits in NGC~6791 and combine this with radial velocity
measurements we can put constraints on the mass and radius of the parent 
star and in turn also constrain the radius of the transiting object 
from the depth of the transit.

In Sect.~2 we describe the observations and in Sect.~3 we discuss
the data reduction. In Sect.~4 we present the calibrated colour-magnitude
diagram. In Sect.~5 we discuss the detected transit candidates. In
Sect.~6 and~7 we present the new variable stars and the 
detached eclipsing binaries. Finally, in Sect.~8 we discuss 
our results and suggest future prospects for planet searches though photometry.

\section{Observations}

The observations for this project were acquired between July 9 and 
17, 2001 at the Nordic Optical Telescope (NOT) by one of us (Bruntt). 
We used the \alfosc\ instrument, which 
gives a field of view of 6\Secc5\,$\times$\,6\Secc5. 
A 2k\,$\times$\,2k thinned Loral CCD with a pixel size of 0\Sec188,
read-out noise 6.5\,$e^{-}$ (rms), and a gain of 2.5 $e^{-}$/ADU 
was used as detector. The time-series observations were made using
$V$ and $I$ filters to be able to distinguish between 
grey and coloured transits.
To first order, planets will cause the same transit depth
regardless of colour, 
while a transit due to a stellar companion or variations due to the 
star itself (spots) will most likely have different amplitudes 
in different filters. The typical exposure times used
were 180\,s and 90\,s for $V$ and $I$ respectively, chosen such that stars above the
turnoff of the cluster at around $V=16$ 
would be at the saturation limit.  In order to reduce the overhead in 
connection with changing filters we grouped the observations as 
five exposures in $I$ followed by three in $V$. The read-out time of the
detector was 70\,s.

A few observations were collected
in the $B$ filter as well, to be used for the construction of the 
colour--magnitude diagram of the cluster.

In order to achieve the highest possible
precision in the differential photometry 
we made small corrective offsets of the 
telescope during the night to keep the stars in the same
position on the detector. 
To be specific, the 1-$\sigma$ RMS scatter of the $x$ and $y$  
positions was 2.8~pixels and the median seeing was 5.3~pixels.

Flat fields were obtained with different rotations of the camera 
(to minimize scattered light effects; see Grundahl \& S\o rensen \cite{grundahl96}) during 
evening and morning twilight on each night. 
A master flat field was constructed by using all flat field images, 
since no significant night-to-night variations were seen.

On each night we observed several standard fields 
from Stetson (\cite{stetson_standard}) in order to provide a good transformation to the standard $BVI$ 
system\footnote{\label{cadc}Available from {\tt http://cadcwww.hia.nrc.ca/cadcbin/ wdb/astrocat/stetson/query/.}} 
(Johnson $BV$ and Cousins $I$).
The details of this can be found in Stetson \ea\ (\cite{stetson03}).
For the remainder of this paper we will refer to the data set from this run as
{\em NOT01}.




The fundamental parameters of \ngcsyv\ are given in Table~\ref{tab:n6791_fund}.

\begin{table*}
\centering
\caption{Fundamental properties of \ngcsyv: right ascension, declination, age, distance modulus,
metallicity, and reddening. The values are taken from Chaboyer \ea\ (\cite{n6791_par}) except the
metallicity which is from Taylor (\cite{taylor}). Note that the quoted errors are 
probably too optimistic and do not include systematic errors (except the error on the metallicity).
\label{tab:n6791_fund}}
\vskip 0.1cm
\begin{tabular}{cccccc}
$\alpha_{2000.0}$ & $\delta_{2000.0}$ & Age [Gyr] & $(m-M)_V$  & [Fe/H] & $E(B-V)$ \\ \hline
19$^{\rm h}$20$^{\rm m}$53$^{\rm s}$ & $+37^{\rm \circ}$46$^{\rm '}$30$^{\rm ''}$ & 
 $8.0(5)$ & 13.37(10) & $+0.30(15)$ & 0.10(2) \\
\end{tabular}
\end{table*}

\subsection{Time-Series Studies of \ngcsyv}


Rucinski \ea\ (\cite{rucinski96}) carried out a campaign on \ngcsyv\ for 14 nights 
in July 1995 to search for detached eclipsing binaries. 
Despite good weather conditions
the noise level in this data set (labelled {\em SMR95} from this point on)
is higher compared to the {\em NOT01} data due to the smaller telescope: 
the apertures are 1\,m for {\em SMR95} and 2.5\,m  for {\em NOT01}.
The {\em SMR95} data is still useful for constraining the 
periods of long-period variables and will be used to search for
detached eclipsing binary stars.
Rucinski \ea\ (\cite{rucinski96}) discovered 11 variable stars (four new ones)
and among them were three detached binaries. 
We refer to Rucinski \ea\ (\cite{rucinski96}) for a list of earlier studies on \ngcsyv.

A more recent study was made by Mochejska \ea\ (\cite{mochejska02}) who have
selected \ngcsyv\ as their first target to search for planetary transits.
They have used the {\code{FLWO}} 1.2\,m telescope at Mt.\ Hopkins, Arizona,
from July 6 to August 1, 2001 (we label this data set {\em MOCH01}), 
\ie, partly overlapping with the {\em NOT01} observations from La Palma.
Unfortunately, the weather conditions 
were quite poor with a median seeing 
in $V$ of 2\Sec1 (in contrast, for {\em NOT01} the median seeing was 1\Sec0).  
Mochejska \ea\ (\cite{mochejska02}) state that the search for transit-like events still 
remains to be done, and that a new campaign on the cluster is being planned.
Mochejska \ea\ (\cite{mochejska03}) have analysed time-series data
spanning six years and report the discovery of seven new long-period and irregular 
variables. Kaluzny (\cite{kaluzny03}) reported the discovery of four new variables
in the field.

In Table~\ref{tab:n6791} we give a summary of the observations of
\ngcsyv\ that we have used.
For the {\em MOCH01} data we have only used the light curves of 
the variable stars from Mochejska \ea\ (\cite{mochejska02}); in other words, we have not 
reduced these data ourselves as we have for {\em NOT01} and {\em SMR95}.
Note that the photometric precision of the {\em NOT01} campaign is superior
to the other telescopes, and only these data (seven nights) were
used to search for transit-like events.

In addition to the data listed in Table~\ref{tab:n6791} several 
images taken outside the center of \ngcsyv\ from {\em NOT01} and
several images from other telescopes 
were reduced with \allframe\ as described by Stetson \ea\ (\cite{stetson03}).

\begin{table*}
\caption{Log of observations of \ngcsyv\ from three campaigns.
The columns contain the adopted abbreviation 
of the campaign, the name of the telescope, the duration
of the campaign, the number of useful nights ($n$), the median seeing in the $V$ filter,
and the number of images obtained in each of the $BVRI$ filters. 
The last column gives information on which reduction
technique we have used. Note that we have not reduced the {\em MOCH01} data set, 
but we have obtained the light curves
of the variable stars published by Mochejska \ea\ (\cite{mochejska02}).
\label{tab:n6791}}
\vskip 0.2cm
\centering
\begin{footnotesize}
\begin{tabular}{llrc|c|cccc|r}
 &  &   &  &  &  \multicolumn{4}{c|}{Number of images} & \multicolumn{1}{c}{Reduction} \\
Campaign & Telescope& Dates UT & $n$ &  [``] & $N_B$ & $N_V$ &$N_R$ & $N_I$ & \multicolumn{1}{c}{Method}\\
\hline
{\em NOT01}  & 2.5\,m NOT & 9/7--17/7 01    &  7 & 1\Sec0  & 58 & 264 &   - & 444 & \allframe/\isis \\
{\em MOCH01} & 1.2m FLWO  & 6/7--1/8 01     & 18 & 2\Sec1  & -  &  36 & 204 &   - & {\em Not Reduced} \\
{\em SMR95}  & 1.0m JKT   & 11/7-24/7 95    & 14 & 1\Sec6  & -  & 601 &   - & 153 & \allframe \\
\end{tabular}
\end{footnotesize}
\end{table*}

\section{Photometry of \ngcsyv}

We have used two different methods for obtaining the differential
time series photometry: (1) the profile fitting (PSF) approach 
viz.\ \daophot/\,\allstar/\,\allframe\ (Stetson \cite{stetson87,allframe}) and (2) 
the image-subtraction method using the \isis\ software (Alard \& Lupton \cite{alard98}).
We shall describe our approach for these two methods below. 

\subsection{{\allframe} Reductions\label{sec:stetson_w}}


The time-series photometry from {\em SMR95}, the new {\em NOT01} data set,
as well as a few frames from other campaigns
were reduced using the \daophot\,/ \allframe\ photometric
software (Stetson \cite{stetson87,allframe}). This is described in 
our companion paper (Stetson \ea\ \cite{stetson03}).
The individual light curves of the variable stars 
(and the transit-like candidates found from the \isis\ light curves)
were derived from these data as follows.

Transformation equations relating the instrumental CCD magnitudes to
standard-system magnitudes and colours, with corrections for atmospheric
extinction, were derived from standard fields observed the same nights
as the \ngcsyv\ observations were obtained.  These transformations must
then be used again in the reverse sense to convert the observed instrumental
magnitudes for the cluster stars to magnitudes on the standard system.
Since these transformations are expressed in terms of standard, not
instrumental colours, initially each given star was assumed to have a 
``typical'' colour.  Then, robust least-squares methodology was used to 
find the standard-system magnitude in each of the
photometric band-passes which was most consistent with all the observed
instrumental magnitudes for that star.  These new standard-system magnitudes
then implied a new, more realistic colour for the star.  These were used for
a new least-squares solution for the best standard-system magnitudes, and
the process was iterated until the derived magnitudes stopped changing.
Because the colour terms in the
transformations are quite small, with absolute values of well under
0.1$\,$mag/mag, this process converges very quickly.  This approach does
have the drawback that at each epoch the instrumental magnitude is
transformed to a standard-system magnitude based upon the {\it
mean\/} colour of each star, averaged over all epochs.  In the case of a constant star,
this is an optimal approach, but in the case of a variable star it could
introduce small systematic errors in the magnitudes observed at the
extremes of the star's colour cycle.  For a variable star, it would be 
better to correct each
instrumental magnitude according to the star's {\it instantaneous\/} colour,
which is possible when light curves are being fitted at the same time as the
photometric observations are being calibrated.  This was done, for
instance, when we used the same software for the {\it Hubble Space
Telescope Key Project on the Extra-galactic Distance Scale\/}, where
template Cepheid light curves were being fitted to variable-star candidates
(see Stetson \cite{stetson_ceph, stetson98}).  In the present case, we had no
preconceived notions as to the forms of individual variable stars' light
curves, so we did not try to fit light curves during the photometric
reductions in order to be able to predict instantaneous colours; 
the extraction of
light curves was done {\it ex post facto\/} after the calibration of the
photometry.  Note that this procedure may bias {\it instantaneous\/}
standard-system magnitudes by amounts not larger than the
colour-transformation terms times half the colour amplitude of a variable
star, which will generally lead to a {\it maximum\/} systematic error less
than 0.01$\,$mag in the instantaneous magnitudes at the epochs of extreme
colour.  Systematic errors in the periods, {\it mean\/} magnitudes, and
mean colours of the variable stars should be negligible.

 \subsection{Difference Imaging with \isis\label{sec:isis6791}}

We have used the difference-imaging software \isis\ version $2.1$ 
developed by Alard \& Lupton (\cite{alard98}). For each of the 
filters $V$ and $I$, we selected the images with the best seeing, 
which were then averaged to 
generate reference images. For each original image a kernel is computed which
describes the variations of the PSF across that image relative to
the reference image. The final step is to convolve the reference image
by this kernel and subsequently subtract the convolved reference image
from each individual image. In principle what will be left in the subtracted
images will be the signal that is {\em intrinsically} different from the
reference image, \eg, variable stars.


One of the advantages of using the difference-image
technique is that the signal from variable stars 
in crowded regions will be less affected by the 
neighbouring stars, \eg, when compared to profile fitting photometry.
Another important point is that variations due to airmass and 
transparency variations are removed to first order 
as a part of the image subtraction.

We tried to use the photometry package of \isis\ on
the subtracted images. Unfortunately, the program produced only meaningless 
output (all stars were constant).
Instead we used the \daophot\ package to extract aperture photometry
in the subtracted image around the position of all stars in the field
(found by using \allstar).
Normally, \daophot\ calculates the magnitude of a star as 
$m = 25 - 2.5 \log (N)$, where $N$ is the number of counts in the
stellar profile.
But since the number of counts in an aperture can be negative 
for the difference images, we had
to add a short piece of code in the original \daophot\ program to account for this. 
In particular, we defined a new ``difference magnitude'' as 
$m_{\rm ISIS} = 25 - 2.5 \log \left[(N + \sqrt{N_0^2+N^2}) / 2\right].$
The zero-point of this magnitude scale is $m_0 = 25 - 2.5 \log(N_0)$ and we
used $N_0 = 1$, \ie, $m = 25.0$ when $N = 0$. 
One can show that the number of counts will be 
$N = N_0 \sinh[{ 0.4 \ln(10) (m_0 - m + 2.5 \log(2)) }]$.

We used a range of increasing aperture sizes which were scaled
with the seeing (\ie, the measured \fwhm) of each image. 
We then defined magnitude intervals with 
a width of 1~magnitude and selected the aperture size to be used
within each interval which gave the lowest noise level.
Large apertures were used for the bright stars and smaller apertures for
fainter stars (the latter will be more
affected by sky background noise). Based on growth curves
constructed for the direct images (Stetson \cite{daogrow}) we then scaled the
counts in each aperture to include all the light. 

The relative magnitudes were then computed as 
$\Delta m = -2.5 \log [(N_{\rm ref} - N_i) / N_{\rm ref}]$, 
where $N_{\rm ref}$ is the counts in the reference image and
$N_i$ is the number of counts in the chosen aperture after applying 
the correction from the growth curve.

\subsection{Repairing the Images}

Any major defects in the images will cause residuals 
in the subtracted images. There are three
major contributors to the defects: cosmic rays, bad columns, and
saturated stars, including charge leakage extending down the image
columns from the brightest saturated stars. 
While cosmic rays are normally rejected as part of the \isis\ reduction
the latter two effects may influence the quality
of the image subtraction. In parti\-cular 
the image subtraction gives large residuals 
around the most saturated stars.
We thus decided to repair the saturated pixels in the cores of
bright stars and the flux from stars which fall on bad columns.
As a result of this procedure, which we will now describe, the residuals 
in the subtracted images 
were still present, but only very close to the saturated star.



A short code to replace pixels on the bad columns and the 
saturated pixels was written in \idl. 
For each image the code examines the pixels around the brightest stars
to look for pixels above the saturation limit (60\,000 ADU). We also
identified stars for which the centroid is within 20 pixels of a bad column.
We then computed an "artificial" image where profiles for these stars 
were inserted using the PSF model found with \daophot\ for each image,
employing for this purpose the positions and instrumental magnitudes found
for these stars by \allstar.
We then replaced the saturated pixels and the signal on the bad columns in 
the observed image with the data from the artificial image
(including the estimated background level and photometric noise).

We stress that when searching the light curves for 
transit-like events and variable stars, we clearly mark the points in the light
curves where the star is close to a bad column or a neighbouring 
saturated star, since the validity of the measured signal in these 
cases is indeed questionable.

\subsection{Photometric Noise in the \isis\ Light Curves}

\begin{figure} 
        \centering
        \includegraphics[width=8.8cm]{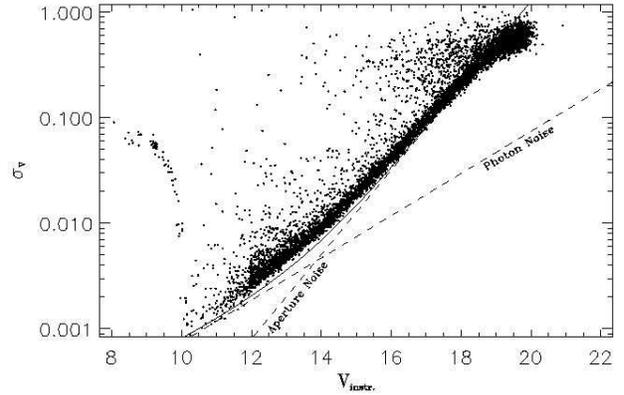} 
        \caption{Measured noise level in $V$ filter (instrumental magnitudes) measured 
over seven nights from the {\em NOT01} data set.
The reductions were made with \isis\ (Alard \& Lupton \cite{alard98}). The dashed lines mark the 
photon noise and the aperture noise while the solid line is the quadratic combination of these two.}
        \label{fig:n6791_noise}
\end{figure}

In Fig.~\ref{fig:n6791_noise} we plot the image-to-image 
noise level of the light curves 
from the {\em NOT01} data set reduced with \isis. 
The dashed lines mark the
contribution to the noise from photon statistics and aperture noise 
(flat field error, sky background determination, and read-out noise) calculated 
using Eq.~(31) in Kjeldsen \& Frandsen (\cite{kjeldsen92}):

\begin{equation}
\sigma^2_{\rm ap} = { {2 \ln 2}\over{W^2\pi e_{\rm ff}} } + { {1}\over{e_{\rm star}} } +
\pi r^2_{\rm ap} { { e_{\rm sky} + \sigma_{\rm ccd}^2 }\over{ ( e_{\rm star} )^2 }}\,,
\end{equation}
where $\sigma^2_{\rm ap}$ is the theoretical total noise level,
$W$ is the FWHM of the stellar profile (in pixels), 
$e_{\rm ff}$ is the number of electrons in the flat field in one pixel,
$e_{\rm star}$ is the total number of electrons from the star,
$r_{\rm ap}$ is the radius of the aperture (in pixels),
$e_{\rm sky}$ is the number of electrons per pixel in the sky background, and
$\sigma_{\rm ccd}$ is the readout noise in electrons.

The solid line gives the combined theoretical noise level.
The brightest stars in the cluster reach a noise level of 1\,mmag per point
while the turnoff stars ($I_{\rm Inst} = 12$) typically have 3\,mmag per point.
Of the 8\,455 stars in our sample, 1\,700 stars have noise $<\,1\,\%$
and for 2\,500 stars the noise is $<\,2\,\%$ per data point.

The fact that the measured noise level is much higher than the 
photon noise level for the faint stars shows that
the combination of image-subtraction and aperture photometry on the
subtracted images is indeed not yet optimal. 

The relative transit depth caused by a close-in giant planet is equal to 
the square of the radius ratio of the planet and the parent star.
For stars below the turnoff region ($I_{\rm Instr} \simeq 12$),
the transit depth of a Jupiter-sized planet will be
of the order 1--3\% (\cf\ Fig.~2 in Mochejska \ea\ \cite{mochejska02}). Thus the photometric noise limit
is well below the limit for detection of transits since we will have several
measurements during a transit which is expected to be of the order 1--3 hours.

\section{The Calibrated Colour-Magnitude Diagram\label{sec:n6791_cmd}}

\begin{figure}
        \hskip -0.5cm \includegraphics[width=9.5cm]{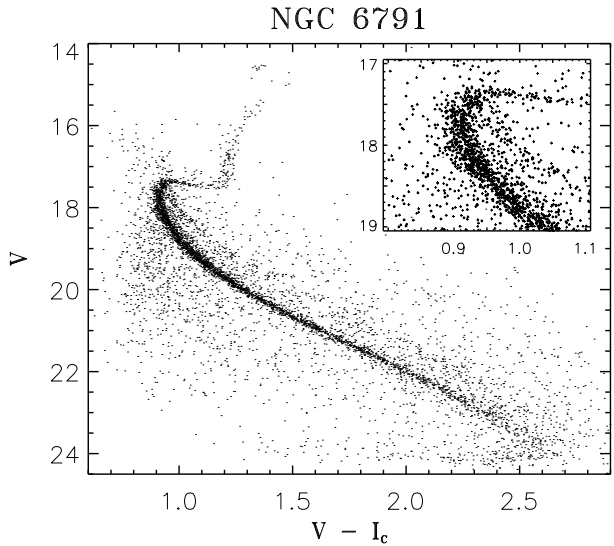}
        \caption{The calibrated colour-magnitude diagram of \ngcsyv.
The inserted plot shows the turnoff region: notice the
narrow main sequence and the binary sequence.}
        \label{fig:cmd6791}
\end{figure}

In Fig.~\ref{fig:cmd6791} we present the colour-magnitude diagram of \ngcsyv.
It is the result of the \allframe\ reduction of 958 $V$ and 737 $I$ frames which 
is described in our companion paper (Stetson \ea\ \cite{stetson03}). 
The magnitudes are calibrated to the standard $V$ and $I$ system.
The formal error on the turnoff stars is around 0.5\,mmag (standard error of the mean) in
both $V$ and $I$ for stars around the turnoff while the calibration errors are
probably not much greater than 2\,mmag.  It should be noted, however, that these are repeatability
estimates for the magnitudes and colours of individual stars contained 
within this data set.  Stetson \ea\ 
present evidence that two stars that appear to have identical photometric
properties with one filter/detector combination can differ by up to 20\,mmag
(standard deviation) when measured with different equipment, 
in these broad bandpasses.  
Nevertheless, the calibrated photometry could be very useful as a standard field
and is available from the 
Canadian Astronomy Data Centre (see footnote on Page~\pageref{cadc}).

\section{The Search for Transit-like Events\label{sec:tingley_w}}

The first seven nights of the {\em NOT01} campaign were used in the
search for transit-like events. The quality of the eighth night was
poor by comparison and neglected for that reason. Moreover, the data
from the {\em SMR95} campaign was also considered to be too poor for
the detection of low-amplitude transits. The {\em NOT01} light curves were searched
for low-amplitude transits using the
matched filter technique, first suggested for this task by 
Jenkins et al.\ (\cite{jenkins96}). 
The decision to use this method to identify transit-like features
is based on the work of Tingley (\cite{tingley03}), where it is shown that the matched
filter performs best. 
The data from {\em NOT01} was of very good quality and
very consistent, with no variation of the signal-to-noise ratio from
night to night.

Our first implementation of this transit-detection algorithm searched
for single transits without using point-to-point noise, which is reasonable
considering the consistency of the data. It assumed a square-well transit
shape and looked for transits with durations of 2, 3, and 4 hours and depths
of 1, 2, and 3\%. A histogram was made of the times of the derived events, to
determine if any particular epochs were over-producing anomalies, which in
theory should be distributed evenly in the case of uncorrelated noise. The
epochs that over-produced events were then removed and the analysis repeated.
This process was performed several times, which significantly reduced the
number of false alarms. Many of the events that remained were also easily
identifiable as false alarms, as they could be explained by the
contamination of a star by a nearby known variable star, a saturated neighboring star, 
or by the star being close to the edge of the CCD or a bad column. 
However, many interesting low-amplitude variable stars were identified, 
including ones which were not clearly identified by the 
Stetson $J$-index (\cf\ Sect.~\ref{sec:n6791_vars}).

The second implementation of the transit-detection algorithm searched for
multiple transits using point-to-point noise. The point-to-point noise
was calculated by choosing a sample of some tens of stars in a relatively
narrow magnitude bin with similar, low signal-to-noise ratios. These values were
then weighted by the the overall signal-to-noise of the individual light
curves. Again using the matched filter with a square-well transit shape,
the light curves were searched for repeating transits with periods ranging
from 0.8 to 6.3 days, durations of 2, 3, 4, and 5 hours, 
and depths of 0.02, 0.01, and 0.005 magnitudes. 
This implementation identified fewer strong events, but fewer false alarms 
and variable stars as well, the latter due to the fact that most of them 
were excluded from the experiment due to their relatively high signal-to-noise ratios. 

\begin{figure}
  \hskip -0.5cm   \includegraphics[width=9.5cm]{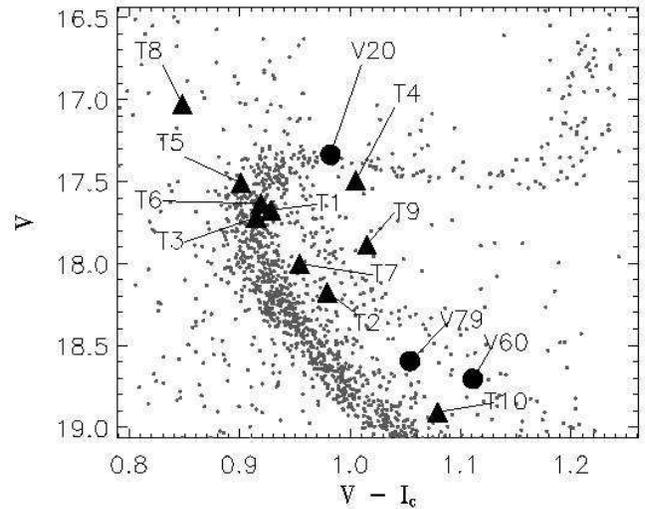}
      \caption{Colour-magnitude diagram of \ngcsyv\ with standard $V$ 
               and $V-I_c$ magnitudes from Stetson \ea\ (\cite{stetson03}). 
               The transit candidates (triangles) 
               and the eclipsing binaries (circles) are marked.
               The stars with the best photometry are also plotted
               to emphasize the location of the main sequence, the turnoff, 
               and the red giant branch.
        \label{fig:cmd_n6791_transits}}
\end{figure} 

\begin{figure*}
   \centering 
   \includegraphics[width=16cm]{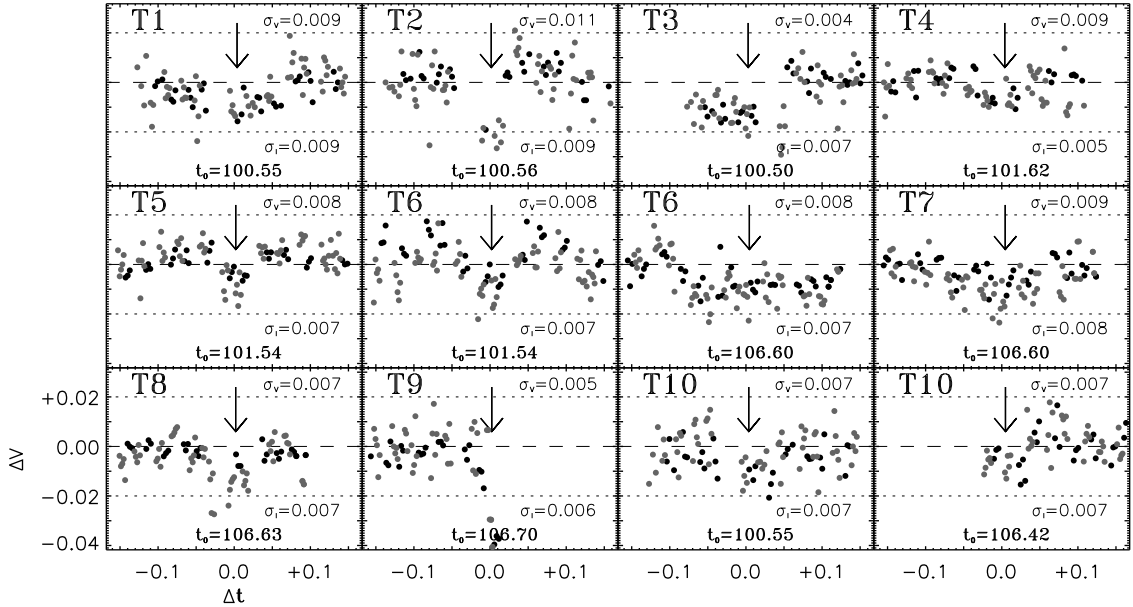} 
      \caption{Light curves of transit candidates.  
               Black and grey symbols are $V$ and $I$ measurements, respectively, and 
               the noise level is given as $\sigma_V$ and $\sigma_I$.
               The time $t_0$ in each panel corresponds to 
               the time HJD $-$ 2\,452\,000 of the tick-mark at 0.0.
               Note that T6 has two transits-like events but of 
               very different duration.\label{fig:planet_transits}}
\end{figure*} 

\subsection{Transit-like Light Curves}

After rejection of obvious false alarms, we have found ten stars with
transit-like events in their light curves.
The positions of these stars in the 
CMD are given in Fig.~\ref{fig:cmd_n6791_transits}
and their properties are given in Table~\ref{tab:n6791_transits}.
We have also plotted the stars with the very best photometry, \ie, 
3\,581 stars having the following quality-indices from \allframe: 
$|sharp| < 0.25$, $|\chi-1| < 0.25$, $\sigma_V < 0.04$, and $\sigma_I < 0.04$.
The $sharp$ index measures how well the widths of the PSF and the observed
star agree, $\chi$ is essentially the standard goodness-of-fit index, 
and $\sigma_V$ and $\sigma_I$ are the estimated RMS errors on the PSF magnitudes.
We refer to Sect.~4.1 of Stetson \ea\ (\cite{stetson03}) for a detailed
discussion of the meanings of these indices.


In Fig.~\ref{fig:planet_transits} we present portions of the light curves for
the ten stars with transit-like light-curve features in the {\em NOT01} data set.
The black points are $V$ measurements while the grey points are $I$ measurements.
The times indicated on the plots as $t_0$ are 
HJD $-$ 2\,452\,000 for the tick-mark corresponding to 0.0. 

We note that the telescope was offset slightly on the first night at the 
time $t_0=100.52$ to the position which was used for the rest 
of the {\em NOT01} campaign.
This time coincides with the candidate transits seen in T1, T2, and T3,
and with the first of two candidate transits in T10.
Similarly, T6, T7, and T8 all have apparent transits around $t_0=106.60$, 
but we note that the durations of these events are generally different.
Finally, the observed transit-like event of T2, is very likely due 
to the star being close to a bad column on the CCD. 

We have only observed single transit-like events except for T6 and T10.  
However, the durations of the two events seen for T6 are very different.
T9 is found to be a long-period low-amplitude variable 
(V80, \cf\ Fig.~\ref{fig:n6791_new}), so quite possibly the dip seen
at $t_0=106.7$ could be the beginning of an eclipse in a binary pair.
This seems plausible from the position 
of T9 in the CMD in Fig.~\ref{fig:cmd_n6791_transits}.
The transit-like event seen for T4 is very odd, since the $I$ measurements
indicate a much longer transit duration than the $V$ measurements.


The most promising cases we have found are T5, T7, and T8, for which
the transit events have a duration of 1--3 hours. Furthermore, the depths 
are around 10\mmag\ in both the $V$ and $I$ filters. These facts are not
in contradiction with the expected transit light curves of 
the short-period giant planets which are known from radial-velocity searches
(Butler \ea\ \cite{butler02}). 

As pointed out by Mall\'{e}n-Ornelas \ea\ (\cite{mallen2003}), the shape 
of a transit caused by a planet will be characterized by steep ingress/egress
slopes and a flat bottom during the transit.
The precision of the present photometry 
is not good enough to see the detailed shape of these shallow events.
We would need further data of higher precision to confirm the transits
and also to detect multiple transits.

We finally note that the age of \ngcsyv\ is $8.0\pm0.5$\,Gyr according 
to Chaboyer \ea\ (\cite{n6791_par}) (although we note that Stetson \ea\ (\cite{stetson03})
suggest the age could be significantly greater).
Hence the expected rotational periods of the stars at the turnoff would
be of the order of weeks, and therefore star spots do not seem like a plausible
explanation for the events we have found. 

To summarize, we cannot claim that what we have observed is indeed caused 
by a giant planet in any of the cases presented in 
Fig.~\ref{fig:planet_transits} and Table~\ref{tab:n6791_transits}.


\begin{table*} 
\caption{We give the name, ID number, 
$\alpha$, $\delta$, $V$, $V-I$, and $B-I$ for the stars with transit-like
light curves. 
All information except the adapted name in the first column 
is from Stetson \ea\ (\cite{stetson03}).
The positions of the stars in the CMD is shown in Fig.~\ref{fig:cmd_n6791_transits}
and the light curves can be seen in Fig.~\ref{fig:planet_transits}.
As discussed in the text, the transits seen for T1--4, T6, and T9 are 
not likely to be due to transiting giant planets (they are marked by a $*$ symbol). 
\label{tab:n6791_transits}}
\centering
\begin{tabular}{lr|cc|ccc}
\multicolumn{1}{c}{Name} & \multicolumn{1}{c}{ID} & $\alpha_{2000.0}$ & $\delta_{2000.0}$ & $V$ & $V-I$ & $B-I$ \\  
\hline
   T1$^*$&$ 4891   $& 19$^h$20$^m$43\secx37&$ +$37$^\circ$47\Secc31\Sec5 & 17.675 &  0.928 &  1.818 \\ 
   T2$^*$&$ 6598   $& 19$^h$20$^m$48\secx65&$ +$37$^\circ$47\Secc41\Sec1 & 18.175 &  0.979 &  1.901 \\ 
   T3$^*$&$ 12616  $& 19$^h$21$^m$07\secx28&$ +$37$^\circ$47\Secc39\Sec9 & 17.718 &  0.915 &  1.809 \\ 
   T4$^*$&$ 9049   $& 19$^h$20$^m$55\secx47&$ +$37$^\circ$44\Secc06\Sec1 & 17.493 &  1.005 &  1.839 \\ 
   T5    &$ 3671   $& 19$^h$20$^m$39\secx07&$ +$37$^\circ$47\Secc26\Sec1 & 17.509 &  0.901 &  1.784 \\ 
   T6$^*$&$ 3567   $& 19$^h$20$^m$38\secx63&$ +$37$^\circ$45\Secc33\Sec3 & 17.635 &  0.919 &  1.811 \\ 
   T7    &$ 3723   $& 19$^h$20$^m$39\secx25&$ +$37$^\circ$45\Secc39\Sec3 & 18.002 &  0.954 &  1.827 \\ 
   T8    &$ 9020   $& 19$^h$20$^m$55\secx40&$ +$37$^\circ$47\Secc23\Sec4 & 17.029 &  0.848 &  1.670 \\ 
   T9$^*$&$ 12390  $& 19$^h$21$^m$06\secx48&$ +$37$^\circ$47\Secc27\Sec8 & 17.886 &  1.015 &  1.947 \\ 
  T10$^*$&$ 8279   $& 19$^h$20$^m$53\secx42&$ +$37$^\circ$48\Secc14\Sec6 & 18.907 &  1.079 &  2.092 \\ 
\end{tabular}
\end{table*}

\begin{table*}
\centering
\begin{footnotesize}
\caption{
The variables in the {\em NOT01} field. The first column contains the variable name, 
where the new variables
are printed with {\em italics} ({\em V79--V100}). Columns 2, 3, and 4 contain ID, right ascension, and 
declination from Stetson \ea\ (\cite{stetson03}). Columns 5 and 6 are the periods and amplitudes (in the $V$ filter) 
we have determined.
The last column gives the variable star type. The positions of the variables in the colour-magnitude
diagram are given in Figs.~\ref{fig:cmd_n6791_transits} and~\ref{fig:cmd_n6791_vars} while the
phased light curves of some of the stars are shown in Figs.~\ref{fig:n6791_known}, \ref{fig:n6791_new},
and~\ref{fig:V09}.
\label{tab:n6791_var}}
\vskip 0.2cm
\begin{tabular}{rr|cc|rrl} 
Name & ID & $\alpha_{2000.0}$ & $\delta_{2000.0}$ & \multicolumn{1}{c}{$P$ [days]} & \multicolumn{1}{c}{$A$ [mag]}   & \multicolumn{1}{c}{Var.\ Type} \\
\hline
   V1    &  6271   & 19$^h$20$^m$47\secx61&$ +$37$^\circ$44\Secc32\Sec0  & 0.2676766 &    0.194 &  EB, close \\  
   V4    &  8576   & 19$^h$20$^m$54\secx22&$ +$37$^\circ$48\Secc23\Sec8  & 0.325667  &    0.046 &  EB, close \\  
   V5    &  5883   & 19$^h$20$^m$46\secx53&$ +$37$^\circ$48\Secc47\Sec9  & 0.312666  &    0.017 &  EB, close \\  
   V6    & 11376   & 19$^h$21$^m$02\secx72&$ +$37$^\circ$48\Secc49\Sec1  & 0.279062  &    0.042 &  EB, close \\
   V9    &  6371   & 19$^h$20$^m$47\secx88&$ +$37$^\circ$46\Secc37\Sec4  & 3.19174   &    0.040$^{a}$ &  EB, close \\  
  V16    & 12695   & 19$^h$21$^m$07\secx59&$ +$37$^\circ$48\Secc09\Sec6  & 4.53      &    0.026 &  EB, close \\  
  V31    & 11307   & 19$^h$21$^m$02\secx47&$ +$37$^\circ$47\Secc09\Sec3  & 3.34154   &    0.015 &  EB, close \\
  V33    &  3886   & 19$^h$20$^m$39\secx81&$ +$37$^\circ$43\Secc54\Sec4  & 2.35944   &    0.040 &  EB, close \\  
  V38    & 11652   & 19$^h$21$^m$03\secx69&$ +$37$^\circ$46\Secc05\Sec9  & 3.845     &    0.023 &  EB, close \\  \hline
  V20    &  8600   & 19$^h$20$^m$54\secx30&$ +$37$^\circ$45\Secc34\Sec7  &14.4698    &    0.3$^{a}$ &  EB, detached \\ 
  V60    & 10746   & 19$^h$21$^m$00\secx70&$ +$37$^\circ$45\Secc45\Sec1  & 7.510/7.1$^{b}$ &    0.4/0.015$^{b}$ &  EB, detached \\ 
{\em V79}&  8943    & 19$^h$20$^m$55\secx21&$ +$37$^\circ$46\Secc39\Sec7  & $\sim$9.0038   &    0.15$^{a}$ &  EB, detached \\ 
{\em V80}$^{c}$
         & 12390   & 19$^h$21$^m$06\secx48&$ +$37$^\circ$47\Secc27\Sec8  & 4.447     & 0.014    &  EB? \\ \hline 
  V14    &  7637   & 19$^h$20$^m$51\secx67&$ +$37$^\circ$45\Secc24\Sec8  & 5.4--5.8  & 0.028    &  Periodic Var. \\ 
  V17    &  3626   & 19$^h$20$^m$38\secx88&$ +$37$^\circ$49\Secc04\Sec6  & 6.42566   & 0.036    &  Periodic Var. \\
  V41    &  7397   & 19$^h$20$^m$50\secx97&$ +$37$^\circ$48\Secc24\Sec8  & 0.4807278 & 0.039    &  Periodic Var. \\
  V53    & 10798   & 19$^h$21$^m$00\secx84&$ +$37$^\circ$44\Secc35\Sec4  & 7.815     & 0.039    &  Periodic Var. \\ 
{\em V81}&  3268   & 19$^h$20$^m$49\secx65&$ +$37$^\circ$48\Secc08\Sec7  & 7.58      & 0.008    &  Periodic Var. \\ 
{\em V82}&  3857   & 19$^h$20$^m$39\secx71&$ +$37$^\circ$47\Secc36\Sec1  & 7.58886   & 0.010    &  Periodic Var. \\ 
{\em V83}&  5848   & 19$^h$20$^m$46\secx40&$ +$37$^\circ$44\Secc14\Sec1  & 7.014     & 0.037    &  Periodic Var. \\ 
{\em V84}&  6305   & 19$^h$20$^m$47\secx71&$ +$37$^\circ$44\Secc58\Sec2  & 1.6333    & 0.03     &  Periodic Var. \\ 
{\em V85}&  7011   & 19$^h$20$^m$49\secx86&$ +$37$^\circ$45\Secc50\Sec9  & 4.100     & 0.007    &  Periodic Var., EB? \\ 
{\em V86}&  7099   & 19$^h$20$^m$50\secx13&$ +$37$^\circ$48\Secc31\Sec7  & 7.39      & 0.011    &  Periodic Var. \\ 
{\em V87}&  8033   & 19$^h$20$^m$52\secx78&$ +$37$^\circ$44\Secc58\Sec8  & 6.971     & 0.006    &  Periodic Var. \\  
{\em V88}&  8088   & 19$^h$20$^m$52\secx91&$ +$37$^\circ$46\Secc36\Sec9  & 6.91      & 0.008    &  Periodic Var. \\ 
{\em V89}&  9432   & 19$^h$20$^m$56\secx64&$ +$37$^\circ$46\Secc36\Sec2  & 5.7617    & 0.028    &  Periodic Var. \\ 
{\em V90}& 10169   & 19$^h$20$^m$58\secx86&$ +$37$^\circ$44\Secc47\Sec2  & 5.71      & 0.008    &  Periodic Var. \\ 
{\em V91}& 10699   & 19$^h$21$^m$00\secx54&$ +$37$^\circ$48\Secc40\Sec7  & 5.13      & 0.011    &  Periodic Var. \\ 
{\em V92}& 11380   & 19$^h$21$^m$02\secx73&$ +$37$^\circ$46\Secc00\Sec8  & 7.24      & 0.008    &  Periodic Var. \\ 
{\em V93}& 12049   & 19$^h$21$^m$05\secx23&$ +$37$^\circ$47\Secc08\Sec5  & 0.94928   & 0.007    &  Periodic Var. \\ \hline 
  B7     & 12652   & 19$^h$21$^m$07\secx40&$ +$37$^\circ$47\Secc56\Sec5  &$\sim$13.3 & 0.15     &  Long period \\  
  V62    & 11475   & 19$^h$21$^m$03\secx06&$ +$37$^\circ$43\Secc51\Sec8  &$\sim$17   & 0.027    &  Long period \\
  V65    &  7920   & 19$^h$20$^m$52\secx47&$ +$37$^\circ$47\Secc30\Sec5  & 8.613     & 0.008    &  Long period \\ 
  V67    & 11645   & 19$^h$21$^m$03\secx67&$ +$37$^\circ$48\Secc03\Sec7  &$\sim$10   & 0.05     &  Long period \\
  V73    &  8108   & 19$^h$20$^m$52\secx97&$ +$37$^\circ$46\Secc52\Sec5  &$\sim$20.9 & 0.02     &  Long Period \\ 
{\em V94}&  4640   & 19$^h$20$^m$42\secx50&$ +$37$^\circ$44\Secc36\Sec9  &$\sim$15.57& 0.01     &  Long period \\ 
{\em V95}&  4805   & 19$^h$20$^m$43\secx05&$ +$37$^\circ$47\Secc32\Sec5  & 9.477     & 0.012    &  Long period \\ 
{\em V96}&  5485   & 19$^h$20$^m$45\secx26&$ +$37$^\circ$45\Secc48\Sec8  & 9.307     & 0.012    &  Long Period \\ 
{\em V97}&  6777   & 19$^h$20$^m$49\secx17&$ +$37$^\circ$49\Secc14\Sec8  & 9.563     & 0.013    &  Long period \\ 
{\em V98}&  9353   & 19$^h$20$^m$56\secx42&$ +$37$^\circ$45\Secc38\Sec4  &$\sim$8.99 & 0.01     &  Long period \\ 
{\em V99}&  9568   & 19$^h$20$^m$57\secx06&$ +$37$^\circ$48\Secc12\Sec2  &$\sim$10   & 0.003    &  Long period \\ 
{\em V100}&11111   & 19$^h$21$^m$01\secx81&$ +$37$^\circ$45\Secc42\Sec0  &12.5215    & 0.01     &  Long period \\ 
\hline
\multicolumn{7}{l}{({\em a}) For the eclipsing binaries V9, V20, and {\em V79} 
the depth of the primary eclipse in the $V$ filter is given.} \\
\multicolumn{7}{l}{({\em b}) V60 is variable with a low amplitude
and a period of $P\sim7.1$\,days while the period estimated from the} \\
\multicolumn{7}{l}{secondary and primary eclipses is $7.510\pm0.005$\,days. The primary eclipse depth is 0.4\,mag in $V$.} \\
\multicolumn{7}{l}{({\em c}) {\em V80} was detected in our search for transit-like events and is also 
given the name {\em T9} (\cf\ Fig.~\ref{fig:planet_transits}).}\\
\multicolumn{7}{l}{Furthermore, {\em V80} is variable with a low amplitude.} \\
\end{tabular}
\end{footnotesize}
\end{table*}


  \begin{figure} 
   \hskip -0.5cm \includegraphics[width=9.5cm]{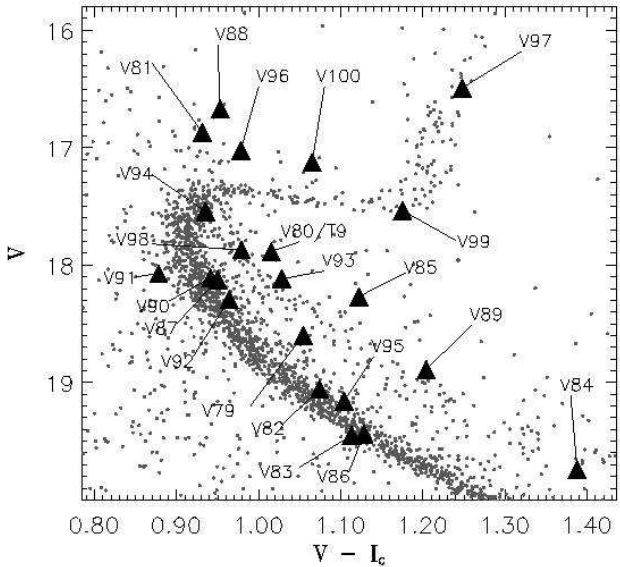}
      \caption{Colour-magnitude diagram of \ngcsyv\ with standard $V$ 
               and $V-I_c$ magnitudes from Stetson \ea\ (\cite{stetson03}).
               The 22 newly discovered variables have been marked.
               Only the stars with the best photometry are plotted
               to emphasize the location of the main sequence, the turnoff, 
               and the red giant branch.
        \label{fig:cmd_n6791_vars}}
   \end{figure}

\section{The Variable Stars in \ngcsyv\label{sec:n6791_vars}}

To search for variable stars we used the Stetson $J$-index (Stetson \cite{stetson_ceph}).
To summarize, the $J$-index is a normalized sum of the deviation of 
each point (or pair of points with small separations in time) in the light curve
compared to the {\em expected} noise level for that star.
To determine the {\em expected} noise level for a given star
we used the mean repeatability of a large number of stars with similar magnitude.

In the following we present the phased light curves of 
some of the variable stars in \ngcsyv. In addition to 20 known variables
in our field we present 22~new variables. The locations in the CMD 
of the 22 new variables are shown in Fig.~\ref{fig:cmd_n6791_vars}.
The new variables all have low amplitudes and long periods. 
This explains why they were not found 
by Rucinski \ea\ (\cite{rucinski96}) or Mochejska \ea\ (\cite{mochejska02}) due to the lower
signal-to-noise ratios in their data sets.

In Table~\ref{tab:n6791_var} we give a list of the variable stars we
have observed. 
The columns in Table~\ref{tab:n6791_var} contain the variable name, 
the ID number, right ascension, declination,
period, the approximate amplitude in the $V$ filter, and the variable type.
The information in columns 2--4 is from Stetson \ea\ (\cite{stetson03}). 

We have determined the period of each star by using the 
method of \sperl\ (Sperl \cite{period98}).
For some of the new long-period variables 
the time series is not long enough or the amplitude is too small
to give a safe estimate of the period.
In these cases we have only given an approximate
period (indicated by the ``$\sim$'' symbol).
Except for some of the long period variables, 
our period estimates agree with previously published results.

\begin{figure*} 
        \centering
        \includegraphics[width=16cm]{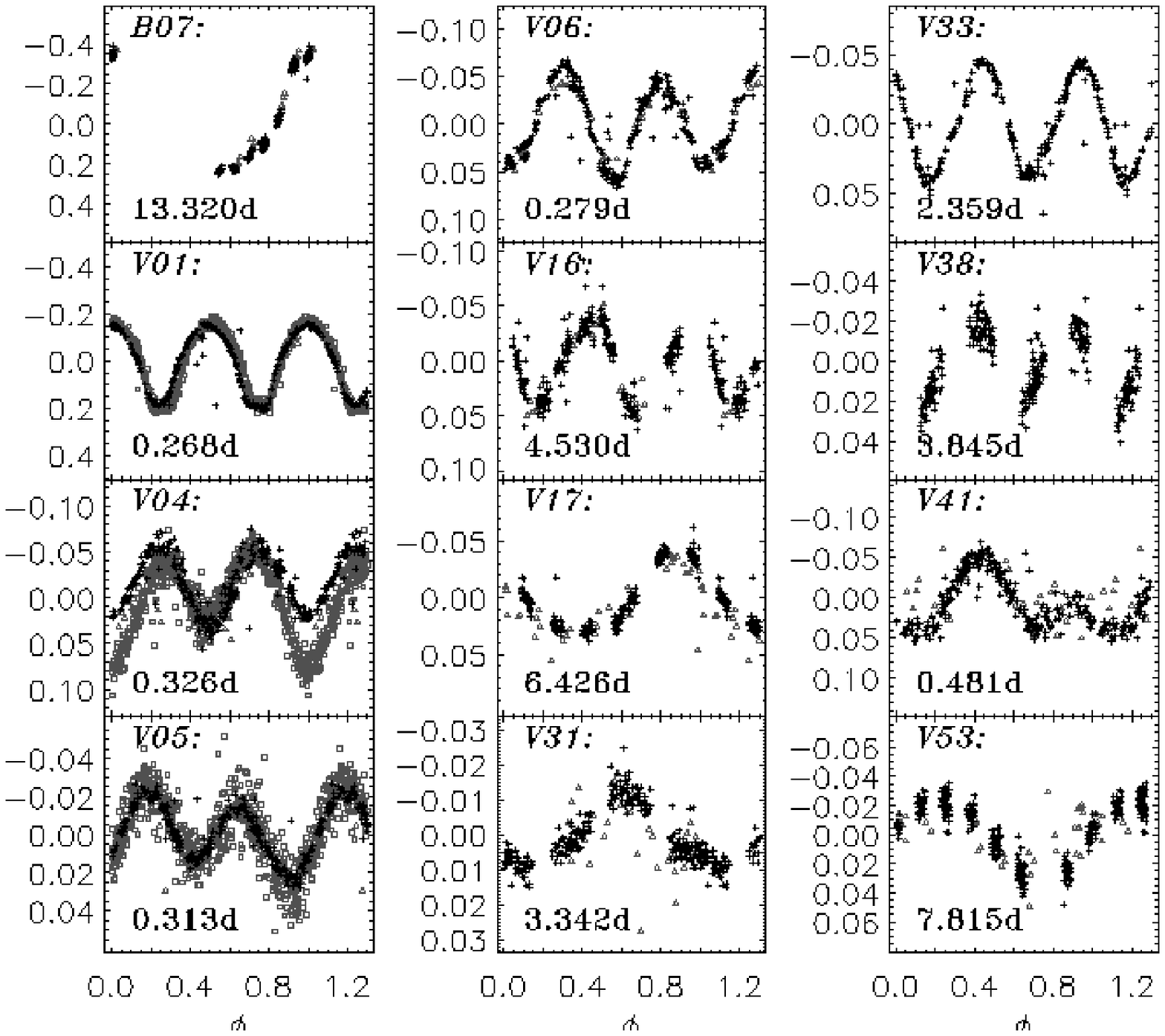} 
        \caption{Phased light curves in the $V$ filter for some of the known 
variables in \ngcsyv. In most panels only the {\em NOT01} data ($+$ symbols) are
plotted due to the much lower noise level compared to 
{\em MOCH01} (triangle symbols) and {\em SMR95} (box symbols). Notice the change
in the light curve of V4.
        \label{fig:n6791_known}}
\end{figure*}

\begin{figure*} 
        \centering
        \includegraphics[width=16cm]{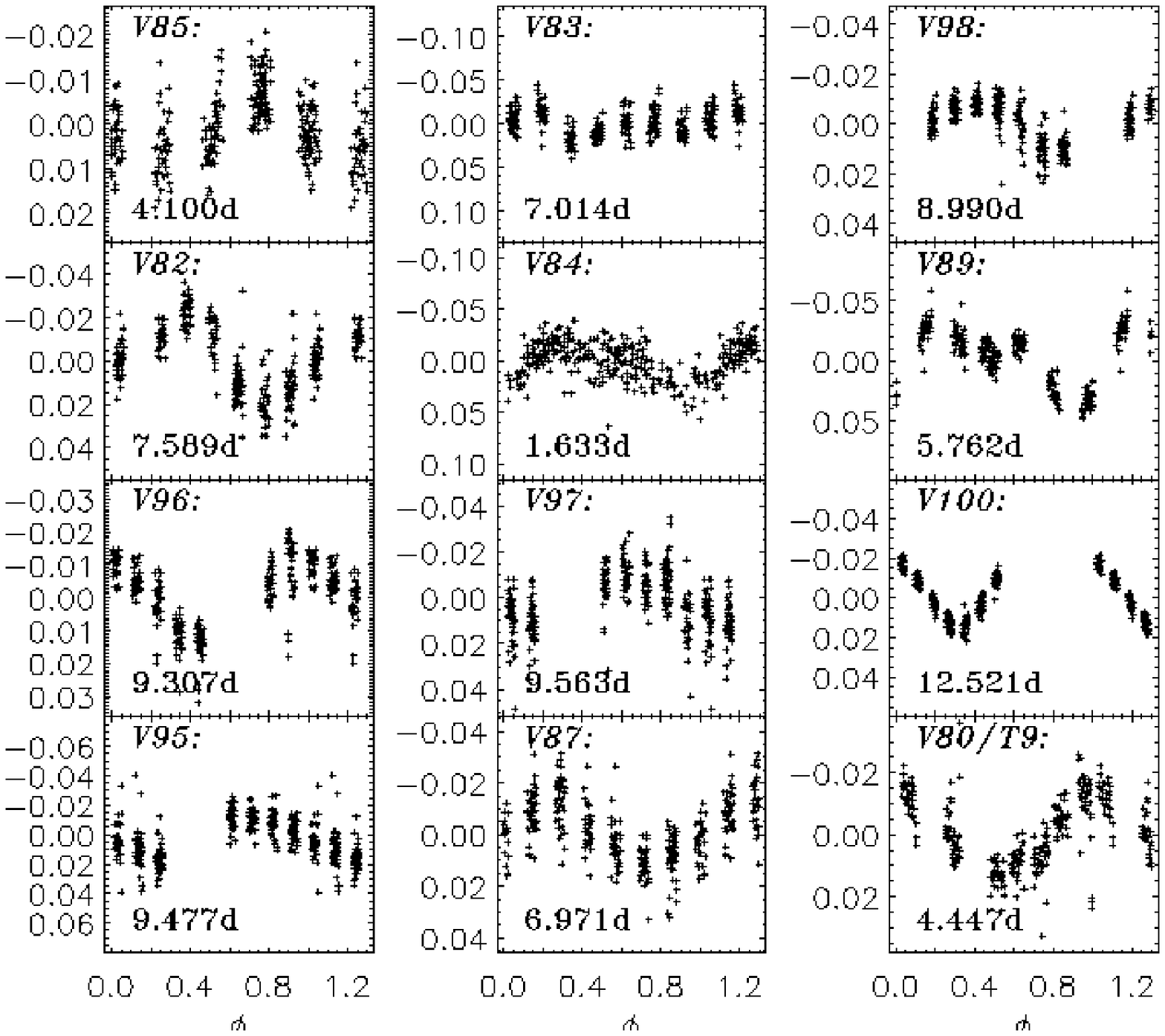} 
        \caption{Phased light curves of some of the new low-amplitude variables. 
Only the {\em NOT01} data are plotted. The ID number is given in the top left 
corner of each panel and correspond to Stetson \ea\ (\cite{stetson03}). The period (in days) is given in the lower
left corner. More details are given in Table~\ref{tab:n6791_var}.}
       \label{fig:n6791_new}
\end{figure*}

In Fig.~\ref{fig:n6791_known} we show the phased light curves of 
some of the known variables 
and in Fig.~\ref{fig:n6791_new} we present some the new variables. 
The ordinate gives the change in magnitude
in the $V$ filter and the phase is on the abscissa.

The {\em MOCH01} and {\em SMR95} data are not plotted in 
Figs.~\ref{fig:n6791_known} and Fig.~\ref{fig:n6791_new} 
if the amplitude of the variable is comparable to the noise 
level in these data sets.

\begin{figure*}
        \centering
        \includegraphics[width=8.8cm]{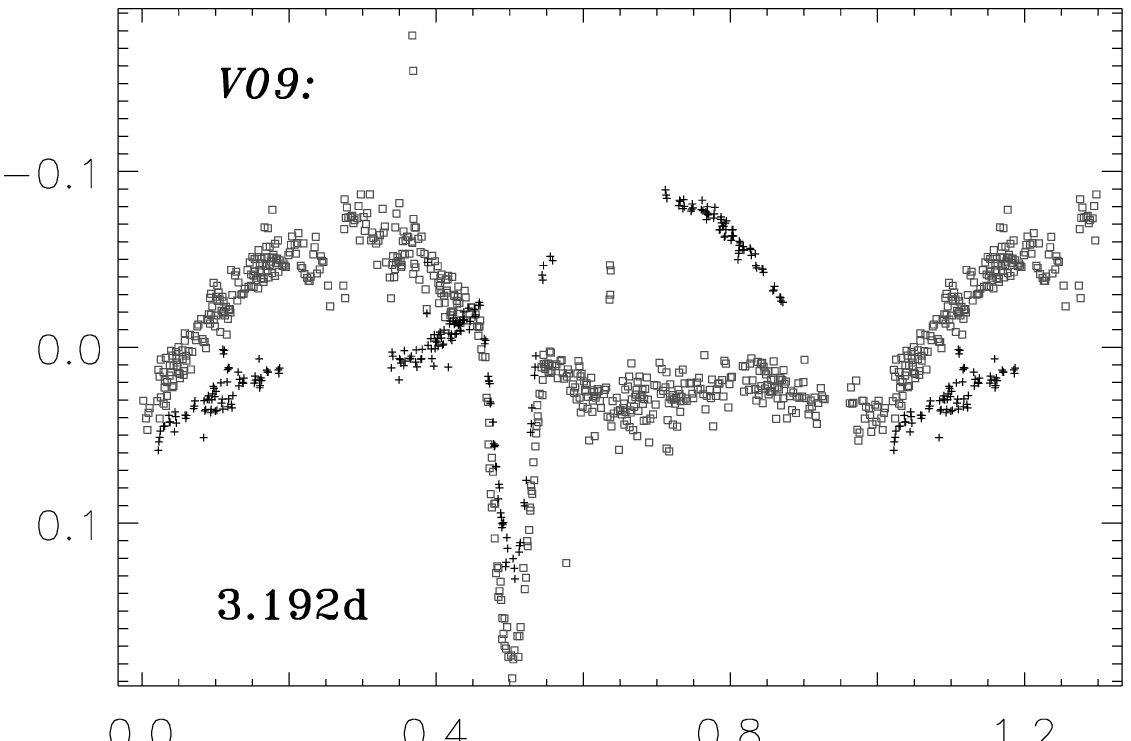} 
        \includegraphics[width=8.8cm]{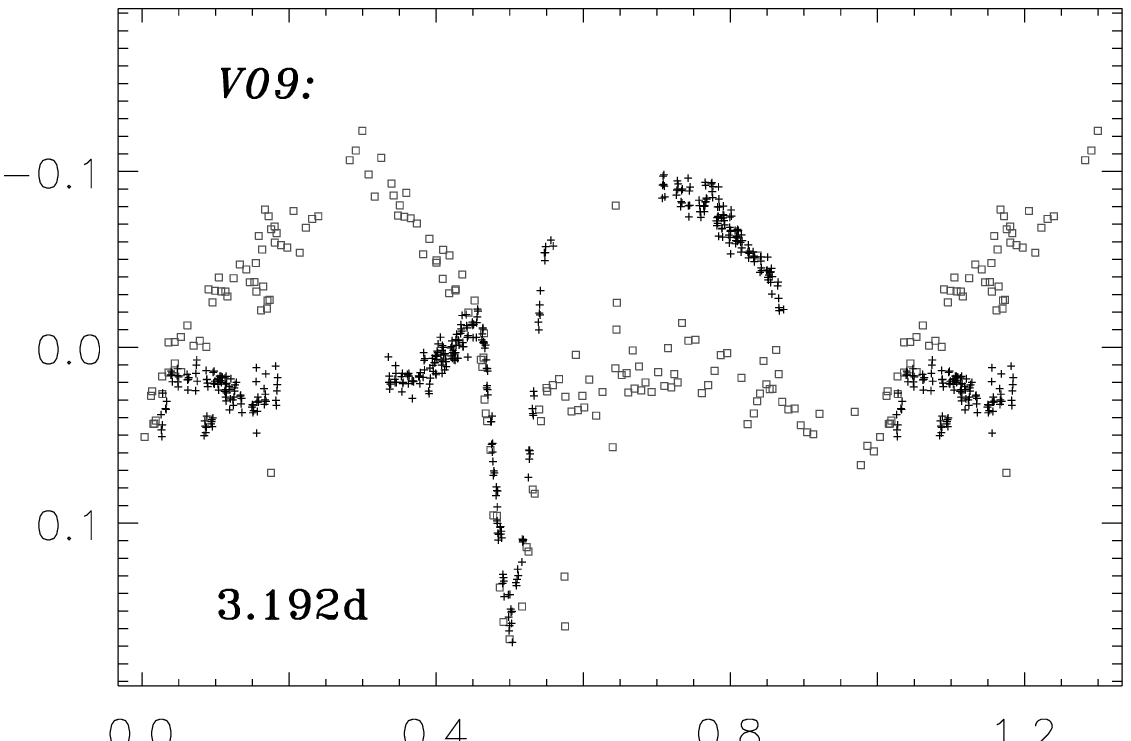} 
        \caption{Phased light of the RS~CV-type variable V9 in \ngcsyv\ in the
$V$ (\lee\ plot) and $I$ filter (\rii\ plot). The black $+$ symbols are data 
from {\em NOT01} while the grey symbols are from either {\em SMR95} (box symbol) or
{\em MOCH01} (filled grey symbols)
The {\em MOCH01} data are not plotted in the \rii\ plot since 
Mochejska \ea\ (\cite{mochejska02}) did not observe in $I$.
        \label{fig:V09}}
\end{figure*}

An interesting eclipsing binary that was previously known is V9, which was observed in all three
data sets. 
It is classified as an
RS Canum Venaticorum type star (RS CVn, see \eg, Hall (\cite{rscv})).
The light curves in the $V$ (\lee\ plot) and $I$ filters are shown in Fig.~\ref{fig:V09} 
where the ``$+$'' symbols are used for the {\em NOT01} data, 
the filled grey symbols are the {\em MOCH01} data, while the squares are the {\em SMR95} data. 
The light curve of V9 is characterized by a deep eclipse which is modulated by
a ``distortion wave'' probably due to dark spots on the surface 
of one of the stars in the binary pair. It is interesting to note 
that while the phase of the eclipse is unchanged the modulation wave 
has shifted phase from 1995 ({\em SMR95}) to 2002 ({\em NOT01} and {\em MOCH01}).
The shape of the modulation wave seems to be unchanged, 
as was also noted by Mochejska \ea\ (\cite{mochejska02}).

We finally note that 
the light curves presented here will be available from the 
Canadian Astronomy Data Centre (Stetson \ea\ \cite{stetson03}).


\section{Three Eclipsing Binary Stars in \ngcsyv\label{sec:binary}}

\begin{figure}
\centering
  \includegraphics[width=8.8cm]{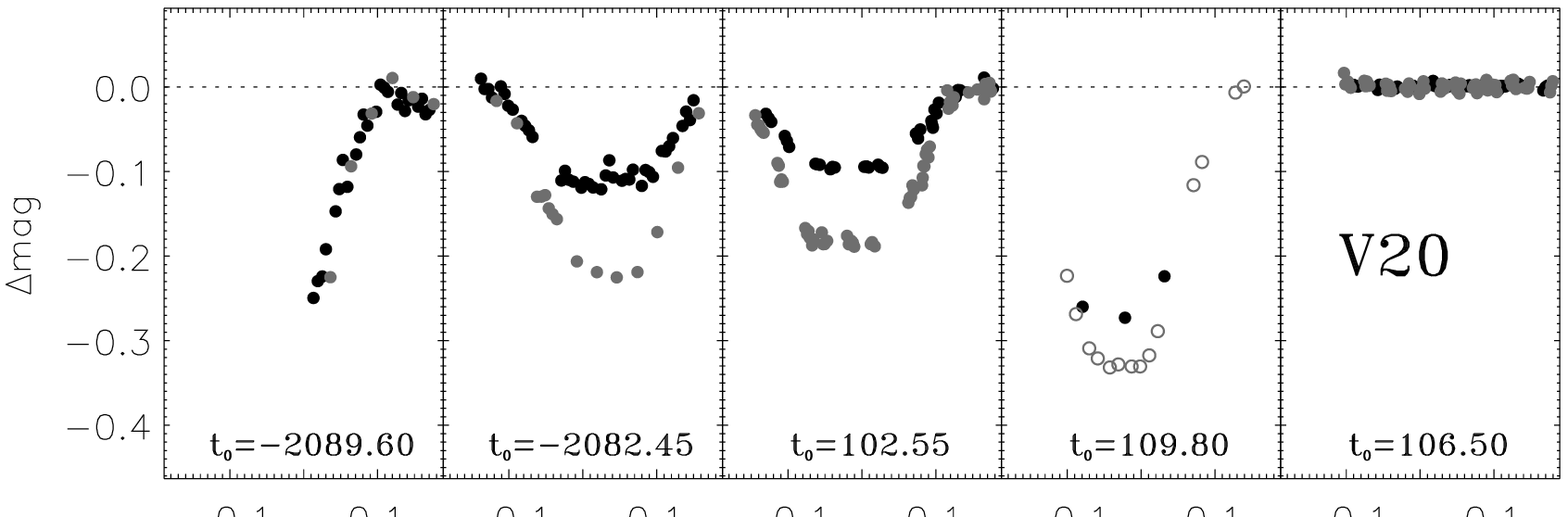}  
  \vskip 0.2cm
  \includegraphics[width=8.8cm]{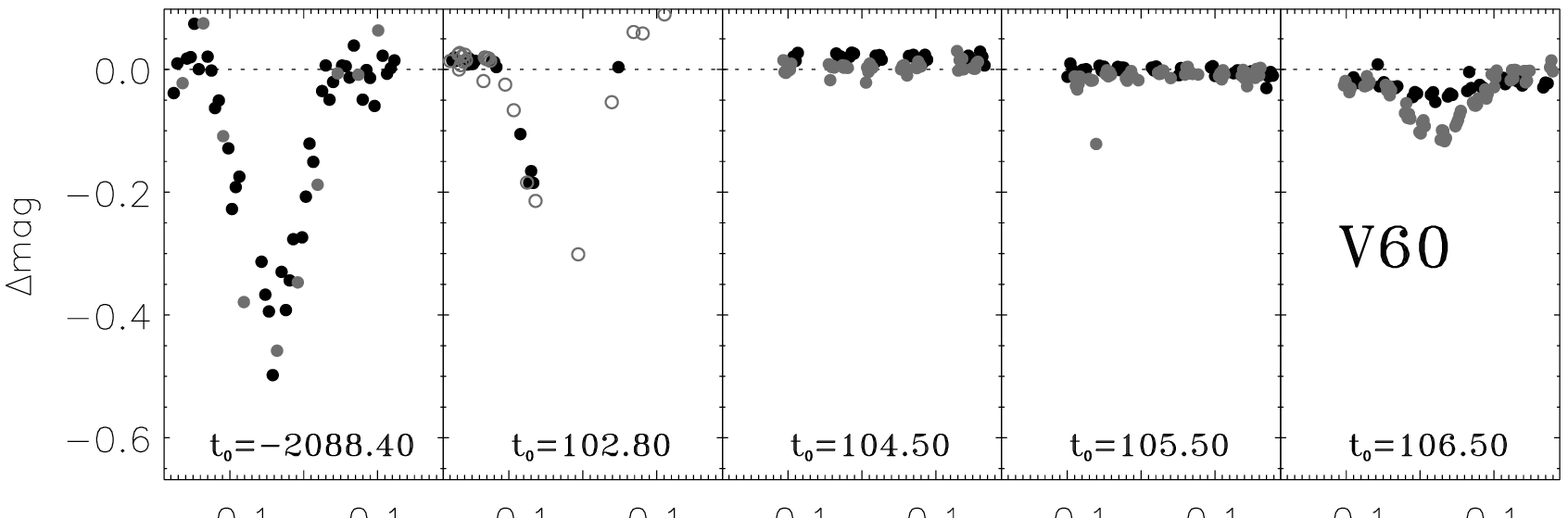}  
  \vskip 0.2cm
  \includegraphics[width=8.8cm]{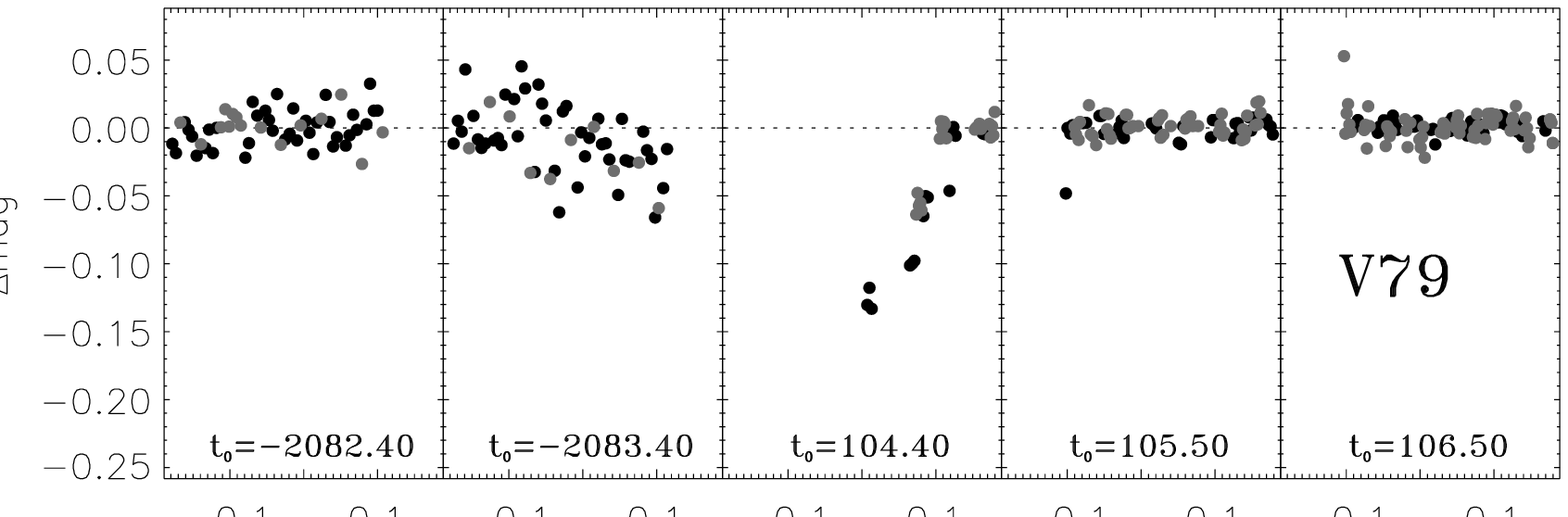}  
  \vskip 0.15cm
      \caption{Light curves of three eclipsing binary stars.
               The time corresponding to 0.0 on the abscissa is HJD $-$ 2\,452\,000 
               and is given at the bottom of each panel as $t_0$. 
               One or more nights {\em without} eclipses are shown for comparison.
               Black and grey symbols are $V$ and $I$ measurements, while the open circles are $R$
               measurements from {\em MOCH01}.
         \label{fig:eclipsing}}
\end{figure}

In Fig.~\ref{fig:eclipsing} we show the light curves of three detached 
eclipsing binaries.
In each panel $t_0$ is the time HJD $-$ 2\,452\,000 for the tick-mark 
corresponding to 0.0. The black and grey points are
$V$ and $I$ measurements, while open circles are $R$ measurements from {\em MOCH01}.

For each of the eclipsing binaries V20, V60, and V79 we show the light 
curves from five nights, \ie, the nights on
which the eclipses occur and one or more nights without eclipses for comparison.

The eclipses of V20 and V60 were noted before by Rucinski (\cite{rucinski96}) and
Mochejska \ea (\cite{mochejska02}), but now for the first time we have enough time coverage 
to be able to constrain the periods. 
From the difference in time between the eclipses 
we find periods of
$P_{\rm bin} = 14.470\pm0.001$ and $7.510\pm0.005$ days for 
V20 and V60, respectively. 
We note that V60 is variable with an amplitude of 0.015\,mag in $V$ 
at a period of $P_{\rm var} \simeq 7.1\pm0.2$ days.


\subsection{The Detached Binary V20}

We have been granted observing time to obtain more detailed photometric coverage 
of the primary and secondary eclipses of V20, and we plan to 
carry out spectroscopic measurements to obtain the
radial velocity curve.
With these observations it will be possible to constrain 
the masses, radii, and luminosities of the binary components of this system.

As will be demonstrated in Sect.~\ref{sec:v20sect},
the primary component of V20 is very close to the turnoff.
From the know\-ledge of the mass of the primary component, 
we expect to be able to determine the age of \ngcsyv\ with 
good accuracy. This can be done from the relation 
between age and mass of the turnoff stars 
which will rely on theoretical isochrones.

This has recently been done for the globular 
cluster \ocen\ by Thompson \ea\ (\cite{thompson01}) and 
Kaluzny \ea\ (\cite{kaluzny02}) for eclipsing stars around 
the turnoff.
Their masses could be determined with an accuracy of just 1\,\%, 
including systematic errors. In this case one of 
the systems is right after the turnoff and Kaluzny \ea\ (\cite{kaluzny02})
were able to constrain the position of the theoretical isochrones 
and in turn estimate the age of \ocen\ to within 0.6 Gyr---at 
least a factor of two better than any 
previous age determination for a globular cluster.



\subsection{Estimating the Parameters of the Detached Binary V20\label{sec:v20sect}}

\begin{figure}
        \centering
        \includegraphics[width=8.8cm]{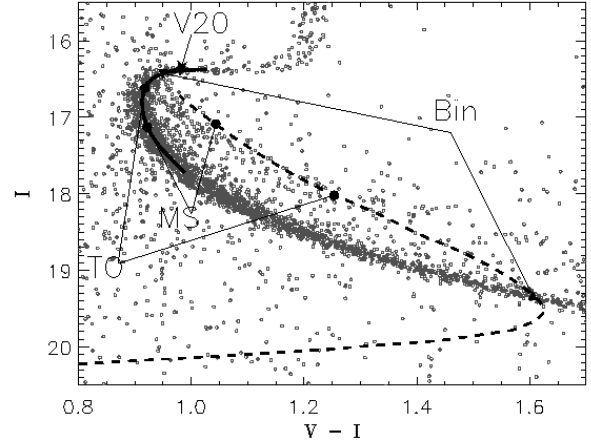} 
        \caption{The $I$ \vs\ $V-I$ colour-magnitude diagram of 
\ngcsyv\ on the standard system. 
The location of the eclipsing binary V20 above the turnoff is shown with
a star symbol. 
To find the possible positions of the two components of the 
binary V20, we assume that the primary star is on the solid black curve
and calculate the corresponding colour and magnitude for the
secondary component which is given as the dashed line. For three points we
have shown the location of the primary and secondary star, \ie, for the
primary star being at the turnoff (TO), further down the main sequence (MS),
and the point where the dashed curve intersects the main sequence (Bin).
        \label{fig:v20_cmd}}
\end{figure}

In Fig.~\ref{fig:v20_cmd} the colour-magnitude diagram of \ngcsyv\ is shown. 
The location of the eclipsing binary V20 is shown with a star symbol 
above the turnoff region. 
In the following we will describe how we have determined the parameters of
the two components that make up the binary V20. 

We have fitted a fiducial locus to the turnoff region 
shown as the solid ``primary'' curve in Fig.~\ref{fig:v20_cmd}, 
and in the following we assume the primary component will be on this curve. 
For each point on the ``primary'' curve we calculate what 
the magnitude and colour of the secondary component should be, while requiring
that the {\em combined} colour and luminosity of the primary and secondary component 
must yield the {\em observed} position of V20. The inferred position of the
secondary component is shown as the dashed ``secondary'' curve.

For three positions on the ``primary'' curve we have marked the 
corresponding position on the ``secondary'' curve: at the turnoff (TO),
on the main sequence (MS), and the point where the main sequence
and the ``secondary'' curve intersect (Bin).
The latter is the best fit 
of the location of the two components making up
the binary V20. Assuming that this inference is correct, their colours are determined to 
within less than $\pm0.01$\,mag; these are given in Table~\ref{tab:v20}.
We will now use these colour estimates to infer the 
properties of the components of V20.


We have used the calibration of \teff\ \vs\ $(V-I)_J$ for main sequence stars with
solar metallicity by Alonso \ea\ (\cite{alonso96}). 
The calibration of Alonso \ea\ (\cite{alonso96})
refer to {\em Johnson} $I_J$ magnitudes, and so we have used the
calibration of Fernie (\cite{fernie83}) to convert from the
{\em Kron-Cousins} system we have used, \ie, $(V-I)_J = -0.005 + 1.273\, (V-I)_C$. 
For the interstellar reddening we use
$E(B-V)=0.105\pm0.025$ (Chaboyer \ea\ \cite{n6791_par}) and using Taylor (\cite{taylor86}) 
we find $E(V-I)=0.13\pm0.03$.
We note that the error given on $E(B-V)$ does not include systematic errors
which could very likely be of the order $\sigma(E(B-V))\simeq0.05$.  However,
the analysis of the colour-magnitude diagram of the cluster by Stetson \ea\ (\cite{stetson03})
suggests a reddening $E(B-V)\sim0.09$, entirely consistent with the
Chaboyer value that we employ.

From these numbers we can determine 
the de-reddened magnitudes and colours and thus estimate the
\teff 's of the two components of V20 to within 200~K, including the systematic
uncertainty in $E(B-V)$. The results are given in Table~\ref{tab:v20}.
We obtain the bolometric correction from
Bessell \ea\ (\cite{bessell}) for atmospheric models with solar metallicity and no overshooting.
We can then determine the relative luminosities of the stars using
$M_{\rm bol; p} - M_{\rm bol; s} = ( V_{\rm p} + BC_{\rm p} ) -  
                                  (V_{\rm s}  + BC_{\rm s} ) = -2.84\pm0.14$
which is equal to $-2.5\log(L_{\rm p}/L_{\rm s})$, and hence 
the luminosity ratio is $L_{\rm p}/L_{\rm s} = 13.7\pm1.8$.
When using $L \propto R^2$\,\teff$^4$ we find a radius ratio 
of $R_{\rm s} / R_{\rm p} = 0.41\pm0.08$.   

We have also estimated the relative luminosities of the binary components by
modelling the observed light curve of V20 following the method outlined in
Clausen et al. (\cite{clausen2003}).
We assumed \teff\,$=\,5\,200$~K for the primary component and
from the modelling of the light curves using only
the light curve in $V$ we find \teff\,$=\,4\,400\pm250$~K 
for the secondary component. The large error in \teff\ is mainly due
to the depth of the secondary eclipse being different by $\simeq 0.02$\,mag 
in the {\em SMR95} and {\em NOT01} data set.
From the light curve modelling we also derive the radius ratio to be 
$R_{\rm s}/R_{\rm p} = 0.55\pm0.1$. Both \teff\ and the radius ratio 
are in good agreement with what we have found by the two methods described here.

Obviously, much more detailed photometric coverage of the primary
and secon\-dary eclipses of V20 are needed to further constrain the
properties of the two components.
As already mentioned, we need spectroscopic
observations of the complete orbit to be able to 
constrain the masses of the components.

\begin{table*}
\centering
\caption{The derived parameters for the two components of the 
binary star V20. We found the colours from Fig.~\ref{fig:v20_cmd}, and
in the text we explain how \teff\ was determined. 
The estimated errors are given in parenthesis.
\label{tab:v20}}
\vskip 0.3cm
\begin{tabular}{c|cccc|l}
Component   & $V$ & $V-I$ & $(V-I)_0$ &  \teff & \multicolumn{1}{c}{$BC_V$} \\ 
\hline
Primary     &  17.38(1) &    0.95(1) & 0.82(6) & 5200(200) & -0.20(5) \\ 
Secondary   &  20.94(1) &    1.61(1) & 1.48(6) & 4070(200) & -0.96(10) \\ 
\end{tabular}
\end{table*}

\section{Conclusion and Future Prospects}

We have described the search for planet transits in the open cluster
\ngcsyv\ from seven nights of observations with the Nordic Optical Telescope. 
The photometric precision is adequate for detecting 
transits by close-in Jupiter-sized planets. We have found ten
stars with transit-like events in the light curves, but 
we argue that most of these are probably not caused by a transiting
giant planet. The observed transit events may be due to 
instrumental effects (bad columns, offset of the telescope) and in one case 
there is evidence that the star (T9) belongs to a binary system.
For the remaining three transit-like events the precision (3\mmag) is not
good enough compared to the very shallow transit depth ($\sim10$\mmag) 
to tell whether the transits are really caused by giant planets, \eg, to see 
whether the light curves are flat during the transit.
To confirm the validity of these three planetary transit candidates 
the observation of multiple transits with higher precision are required.

In July 2002 Piotto, Stetson \ea\ have carried out a 
more ambitious multi-site campaign on \ngcsyv\ to search for planet transits. 
We have obtained observations on eight at least partially clear
nights of a ten-night run on the \cfht\ 3.6\,m telescope 
in Hawaii during 2002 July 2--12. In the same period of time
observations from Mexico and Italy were also carried out.
The weather was not optimal during the 
multi-site campaign on \ngcsyv\ in 2002, but 
we have a higher signal-to-noise ratio compared to the {\em NOT01} data and 
also a higher duty cycle over 10~nights.
This will improve the statistics for detecting multiple transits.
The reduction of data from the 2002 campaign data is now under way.

In addition to our own data from {\em NOT01}
we have used data from two previous photometric campaigns.
The photometric precision of these other data sets is not
high enough to confirm the shallow transits we have found.
We have used the complete data set to redetermine the
periods of the 20 known variables in the {\em NOT01} field.
Furthermore, we have discovered 22 new long-period
stars with low amplitudes. 

We have determined the periods of two detached eclipsing binaries, V20 and V60.
From the position of V20 in the colour-magnitude diagram we are able 
to constrain the temperatures of the binary components. This is in good
agreement with results from the modelling of the primary and secondary eclipses.
We find that the primary component is located just after the turnoff, while
the secondary component is a much fainter main sequence star.
We plan to make thorough photometric and spectroscopic observations of V20
to constrain the masses of the binary components. From the mass of the primary star
we can determine the age of \ngcsyv\ from a comparison with the 
calculated turnoff mass for isochrones of different ages.
This will give an age estimate independent of interstellar reddening.






\begin{thebibliography}{}

   \bibitem[1998]{alard98} Alard, C., \& Lupton, R.\ H., 
     1998, ApJ, 503, 325 
   \bibitem[1996]{alonso96} {Alonso}, A., {Arribas}, S., \& {Martinez-Roger}, C.,
     1996, A\&A, 313, 873 
   \bibitem[1998]{bessell} {Bessell}, M.~S., {Castelli}, F., \& {Plez}, B.,
     1998, A\&A, 333, 231 
   \bibitem[2002]{stepss} {Burke}, C. J., {DePoy}, D. L., {Gaudi}, B. S.,
                  {Marshall}, J. L., \& {Pogge}, R. W., 
     2002, {astro-ph/0208305}
   \bibitem[2000]{burrows2000} Burrows, A., Guillot, T., Hubbard, W. B. \ea, 
     2000, ApJ, 534, 97L
   \bibitem[2002]{butler02} Butler, R. P., Marcy, G. W., Fischer, D. A. \ea,
     2002, Ed., A.\ Penny \ea, IAU Symp.\ 202, ASP Conf.\ Ser., {\em in press}
   \bibitem[1999]{n6791_par} Chaboyer, B., Green, E.~M., \& Liebert, J.,
     1999, AJ, 117, 1360
   \bibitem[2000]{charbonneau00} {Charbonneau}, D., {Brown}, T.~M., {Latham}, D.~W., 
        \& {Mayor}, M.,
     2000, ApJL, 529, 45 
   \bibitem[2003]{clausen2003} Clausen, J.~V., Storm, J., Larsen, S.~S., \& Gim\'{e}nez, A.,
     2003, A\&A, 402, 509  
   \bibitem[2002]{planet_hyades} Cochran, W.~D., {Hatzes}, A.~P., \& {Paulson}, D.~B., 
     2002, AJ, 124, 565
   \bibitem[2002]{dreizler} Dreizler, S., Rauch, T., Hauschildt, P. \ea,
     2002, A\&A, 391, L17 
   \bibitem[1983]{fernie83} {Fernie}, J.~D.,
     1983, PASP, 95, 782 
   \bibitem[2001]{gonzalez01} Gonzalez, G, Laws, C., Tyagi, S., \& Reddy, B.\ E., 
     2001, AJ, 121, 432
   \bibitem[1996]{grundahl96} Grundahl, F., \& S\o rensen, A.\ N.,
     1996, A\&AS, 116, 367
   \bibitem[1972]{rscv} {Hall}, D.~S.,
     1972, PASP, 84, 323 
   \bibitem[1996]{jenkins96} Jenkins, J., Doyle, L., \& Culler, D., 
     1996, Icarus, 119, 244
   \bibitem[2003]{kaluzny03} Kaluzny, J., 
     2003, Acta Astron., 53, 51              
   \bibitem[2002]{kaluzny02} Kaluzny, J., Thompson, I., Krzeminski, W. \ea,
     2002, ASP Conf.\ Proc. 265, 155 
   \bibitem[1992]{kjeldsen92} {Kjeldsen}, H., \& {Frandsen}, S., 
     1992, PASP, 104, 413
   \bibitem[2000]{laughlin00} Laughlin, G., 
     2000, ApJ, 545, 1064 
   \bibitem[2003]{mallen2003} Mall\'{e}n-Ornelas, G., Seager, S., Yee, H.\ K.\ C. \ea, 
     2003, ApJ, 582, 1123   
   \bibitem[1995]{mayor95} {Mayor}, M., \& {Queloz}, D.,
     1995, Nature, 378, 355    
   \bibitem[2002]{mochejska02} Mochejska, B.\ J., Stanek, K.\ Z., Sasselov, D.\ D., 
                  \& Szentgyorgyi, A.\ H.,
     2002, AJ, 123, 3460   
   \bibitem[2003]{mochejska03} Mochejska, B.\ J., Stanek, K.\ Z., \& Kaluzny, J.,
     2003, AJ, 125, 3175
   \bibitem[1996]{rocha96} Rocha-Pinto, H.\ J., \& Maciel, W.\ J., 
     1996, MNRAS, 279, 447     
   \bibitem[1996]{rucinski96} Rucinski, S.\ M., Kaluzny, J., \& Hilditch, R.\ W., 
     1996, MNRAS, 282, 705     
   \bibitem[1998]{salaris98} Salaris, M., \& Weiss, A., 
     1998, A\&A, 335, 943
   \bibitem[1998]{period98} Sperl, M., 
     1998, {Comm.\ in Asteroseismology}, 111, 1
   \bibitem[1987]{stetson87} Stetson, P.\ B.,
     1987, PASP, 99, 191
   \bibitem[1990]{daogrow} {Stetson}, P.~B., 
     1990, PASP, 102, 932  
   \bibitem[1994]{allframe} Stetson, P.\ B.,
     1994, PASP, 106, 250
   \bibitem[1996]{stetson_ceph} Stetson, P.~B.,
     1996, PASP, 108, 851 
   \bibitem[1998]{stetson98} Stetson, P.~B., Saha, A., Ferrarese, L. \ea, 
     1998, ApJ, 508, 491
   \bibitem[2000]{stetson_standard} Stetson, P.~B.,
     2000, PASP, 112, 925
   \bibitem[2003]{stetson03} Stetson, P.~B., Bruntt, H., \& Grundahl, F., 
     2003, PASP, 115, 413
   \bibitem[1986]{taylor86} Taylor, B.~J.,
     1986, ApJS, 60, 577 
   \bibitem[2001]{taylor} Taylor, B.\ J., 
     2001, A\&A, 377, 473
   \bibitem[2001]{thompson01} Thompson, I. B., Kaluzny, J., Pych, W. \ea, 
     2001, AJ, 121, 3089
   \bibitem[2003]{tingley03} Tingley, B.,
     2003, A\&A, 403, 329         
   \bibitem[2002a]{udalski02a} Udalski, A., Paczynski, B., Zebrun, K. \ea, 
     2002a, Acta Astron., 52, 1
   \bibitem[2002b]{udalski02b} Udalski, A., Zebrun, K., Szymanski, M. \ea, 
     2002b, Acta Astron., 52, 115
   \bibitem[2002]{ogle2} Wozniak, P.~R., Udalski, A., Szymanski, M. \ea, 
     2002, Acta Astron., 52, 129   

\end{thebibliography}
\end{document}